\documentclass[prb,aps,twocolumn,showpacs,nobibnotes,epsf]{revtex4}

\usepackage{graphicx}
\usepackage{dcolumn}
\usepackage{bm}
\usepackage{SIunits}

\begin{document}
\title{Magnetic field induced spin-flop transition in Na$_x$CoO$_2$ (0.5$<$x$<$0.55)}
\author{T. Wu, D. F. Fang, G. Y. Wang, L. Zhao, G. Wu, X. G. Luo, C. H. Wang}
\author{X. H. Chen}
\altaffiliation{Corresponding author} \email{chenxh@ustc.edu.cn}
\affiliation{Hefei National Laboratory for Physical Science at
Microscale and Department of Physics, University of Science and
Technology of China, Hefei, Anhui 230026, People's Republic of
China\\}

\begin{abstract}
The isothermal magnetoresistance (MR) with magnetic field (H)
parallel to and perpendicular to ab plane is systematically
studied on the single crystal Na$_{0.52}$CoO$_2$ with charge
ordering at $\sim 50$ K and an in-plane ferromagnetism below 25 K.
The isothermal MR behavior with H $\parallel$ ab plane and H
$\perp$ ab plane is quite different. When H $\parallel$ ab plane,
the MR is always negative and the in-plane ferromagnetic behavior
is enhanced. While the MR with H $\perp$ ab plane changes from
negative to positive with decreasing temperature or increasing H,
and the in-plane ferromagnetic behavior is suppressed. A striking
feature is that the MR with H $\perp$ ab plane shows a hysteresis
behavior below 25 K, which is absent for the case of H $\parallel$
ab plane. These results provide strong evidence for a spin-flop
transition of small moments of Co$^{3.5-\delta}$ sites induced by
H $\perp$ ab plane, leading to a metamagnetic transition for small
moments of Co$^{3.5-\delta}$ sites. These complex magnetism
suggests an unconventional superconductivity in Na$_x$CoO$_2$
system because the Na$_x$CoO$_2$ around x=0.5 is considered to be
the parent compound of superconductivity.

\end{abstract}

\pacs{71.27.+a, 74.70.-b,75.25.+z}

\vskip 300 pt

\maketitle

  The frustrated spin system Na$_x$CoO$_2$ has become a focus in
research due to the discovery of superconductivity with T$_c$
$\sim$ 5 K in Na$_{0.35}$CoO$_2$$\cdot$1.3H$_2$O\cite{Takada}. It
is believed that strong electronic correlations and the possible
novel magnetic ground-states due to geometrical frustration play
an important role in the physics, and a rich phase diagram has
been reported as a function of the Na concentration\cite{Foo}. A
charge-ordering state is found at x=0.5 and novel magnetism is
reported for charge-ordering Na$_{0.5}$CoO$_2$\cite{Foo, Yokoi,
Gasparovic}. Recently, the charge-ordering Na$_{0.5}$CoO$_2$ is
considered to be the parent compound of superconductivity in this
system because the valence of superconductive
Na$_{0.35}$CoO$_2$$\cdot$1.3H$_2$O is found close to
+3.5\cite{Takada2, Barnes, Sakurai}. The superconductivity with
the highest T$_c$ was observed in the vicinity of a magnetic phase
by many groups\cite{Sakurai, Milne, Sakurai2, Ihara}. The
understanding of the novel magnetism is believed to shed light on
the mechanism of superconductivity.

\begin{figure}[h]
\centering
\includegraphics[width=9cm]{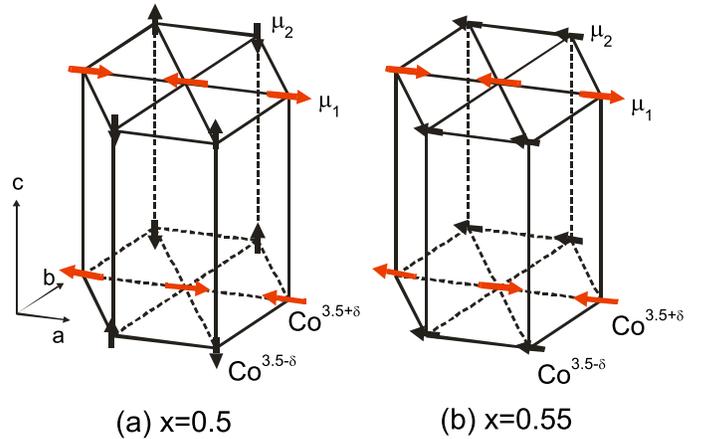}
\caption{Magnetic structure for Na$_x$CoO$_2$: (a) x=0.5 from Ref.
[3, 11]; (b) x=0.55 from Ref. [12].} \label{fig1}
\end{figure}

\begin{figure*}[t]
\centering
\includegraphics[width=18cm]{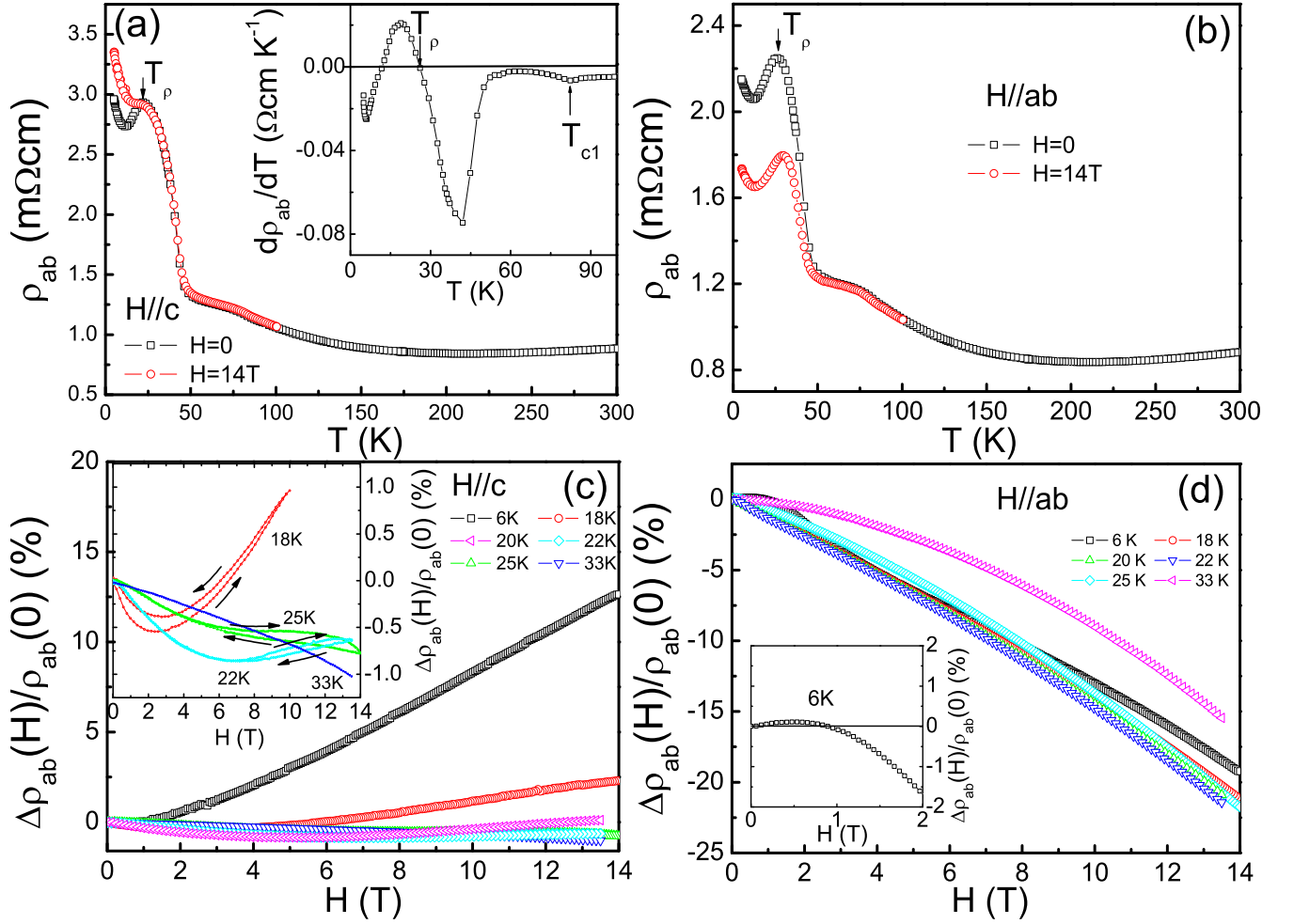}
\caption{(color online). Temperature dependence of in-plane
resistivity under different fields with (a) H $\perp$ Co-O plane
and (b) H $\parallel$ Co-O plane; isothermal magnetoresistance
with (c) H $\perp$ Co-O plane and (d) H $\parallel$ Co-O plane.
The inset in (a) and (c) shows d$\rho$$_{ab}$/dT with H=0 and the
isothermal MR hysteresis, respectively. } \label{fig2}
\end{figure*}

NMR and neutron diffraction studies have given spin structure for
the charge ordering $Na_{0.5}CoO_2$.\cite{Yokoi,Gasparovic} They
reported that there exist two kinds of Co sites with large and
small magnetic moments in $Na_{0.5}CoO_2$. The large moments of
Co$^{3.5+\delta}$ sites align antiferromagnetically at $T_{c1}\sim
87$ K with spin direction within ab plane, while the small
magnetic moments  Co$^{3.5-\delta}$ sites align along the
direction parallel to the c-axis. It cannot be distinguished if
the in-plane spin correlation of the small moment sites is
ferromagnetic or anitferromagnetic. Recently, our group \emph{et
al.}\cite{Wang2} found a six-fold symmetry in angular dependent
in-plane magnetoresistance below a certain temperature (T$_\rho$)
for charge-ordering Na$_{0.34}$(H$_3$O)$_{0.15}$CoO$_2$ sample. It
gives a definite evidence that the small moments of
Co$^{3.5-\delta}$ sites align antiferromagnetically with spin
direction along c axis. The magnetic structure for $Na_{0.5}CoO_2$
is shown in Fig.1(a). Subsequently, Wang \emph{et al.} have found
in-plane ferromagnetism below 17 K in Na$_{0.55}$CoO$_2$ which
enriches the magnetic phase diagram around x=0.5\cite{Wang}. We
found that the small moments of Co$^{3.5-\delta}$ sites align
ferromagnetically in ab plane, and the large moments of
Co$^{3.5+\delta}$ sites still align antiferromagnetically in ab
plane. The magnetic structure proposed by Wang \emph{et al.} is
shown in Fig. 1(b). This result suggests that a spin-flop
transition of the small moments of Co$^{3.5-\delta}$ sites takes
place with slight change of Na content around x=0.5, and the
magnetic coupling of small moments of Co$^{3.5-\delta}$ sites
changes from antiferromagnetic for x=0.5 to ferromagnetic for
x=0.55. Further investigation of the evolution of magnetism in
this region is needed to understand these intriguing magnetic
properties, which is helpful to understand the mechanism of
superconductivity and novel magnetic ground states in this
frustrated spin system.

\begin{figure*}[t]
\centering
\includegraphics[width=18cm]{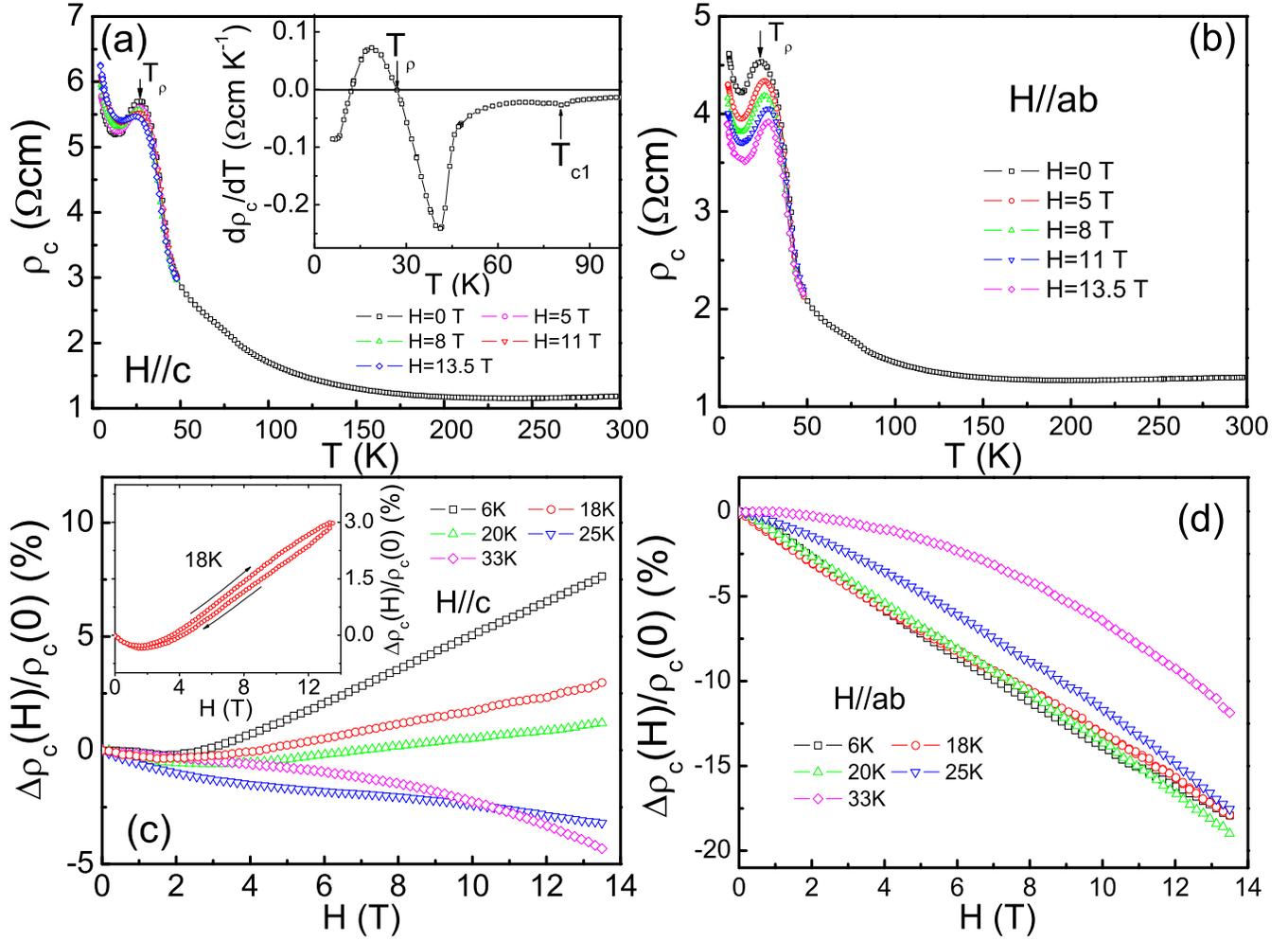}
\caption{(color online). Temperature dependence of out-of-plane
resistivity under different fields with (a) H $\perp$ Co-O plane
and (b) H $\parallel$ Co-O plane; isothermal magnetoresistance
with (c) H $\perp$ Co-O plane and (d) H $\parallel$ Co-O plane.
The inset in (a) and (c) shows d$\rho$$_{c}$/dT with H=0 and the
isothermal MR hysteresis, respectively. } \label{fig3}
\end{figure*}

Magnetoresistance (MR) provides insight into the coupling between
charge and background magnetism. This is particularly valuable
because, as shown in previous work\cite{Wang2}, small magnetic
moments order in the background of magnetic ordering with large
magnetic moments, and such ordering is difficult to detect by
magnetization measurement. In this paper, we systematically study
MR for the Na$_x$CoO$_2$ crystals with x between 0.5 and 0.55. The
MR is defined by
MR=$\frac{\rho(H)-\rho(0)}{\rho(0)}=\frac{\Delta\rho(H)}{\rho(0)}$.
An in-plane ferromagnetism is observed below 25 K. Isothermal MR
shows quite different behaviors between H $\perp$ ab plane and H
$\parallel$ ab plane. A magnetic field induced spin-flop
transition of small moments of Co$^{3.5-\delta}$ sites is proposed
to explain these complex phenomena.

High quality single crystals Na$_{0.7}$CoO$_2$ were grown using
the flux method. The typical dimension is about
2$\times$1.5$\times$0.01 mm$^{3}$ with the shortest dimension
along the c axis. The Na$_{0.52}$CoO$_2$ sample is prepared by
sodium deintercalation of the Na$_{0.7}$CoO$_2$ singe crystals.
The procedure is similar to the way to prepare the
Na$_{0.55}$CoO$_2$ sample; the difference is that a longer
reaction time is needed for the same concentration compared to the
preparation of Na$_{0.55}$CoO$_2$. About 5 mg of Na$_{0.7}$CoO$_2$
single crystals were immersed in the sealed conical flask with 5
ml 0.15 M solution of I$_{2}$ in acetonitrile at room temperature
for about 40 hours. The x values of the samples were estimated by
the lattice parameter c determined by X-ray diffraction. The
resistance was measured by an AC resistance bridge (LR-700, Linear
Research). A superconducting magnet system (Oxford Instruments)
was used to achieve magnetic field up to 14 T. Temperature
measurement used a magnetic field insensitive temperature sensor
(CERNOX, Lakeshore Cryotronics). It should be addressed that all
results discussed as follow are well reproducible.

Fig. 2 shows the in-plane resistivity and magnetoresistance for H
$\parallel$ ab plane and H $\perp$ ab plane, respectively. As
shown in Fig. 2(a), the zero-field resistivity shows charge
ordering behavior at $\sim 50$ K and a kink at 25 K, being similar
behavior as that of the in-plane ferromagnetic
Na$_{0.55}$CoO$_2$\cite{Wang}. The transition of the slope
d$\rho$/dT from negative to positive around 25 K (T$_\rho$) is
believed to arise from the in-plane ferromagnetic ordering for the
small moments of Co$^{3.5-\delta}$ sites\cite{Wang}. A rapid
upturn around 50 K should be due to the charge ordering as that in
Na$_{0.5}$CoO$_2$\cite{Foo}. In addition, another anomaly in slope
d$\rho$/dT is observed at 83 K which corresponds to the
antiferromagnetic ordering temperature (T$_{c1}$) as the case of
Na$_{0.5}$CoO$_2$\cite{Foo}. The temperature in Na$_{0.55}$CoO$_2$
and Na$_{0.50}$CoO$_2$ is 77 K and 87 K, respectively. It further
indicates that the Na content of the current sample is between
x=0.5 and x=0.55, consistent with single crystal XRD result. In
Fig. 2(a) and (b), resistivity under 14 T shows different behavior
below T$_\rho$ with different field direction. When H $\perp$ ab
plane, T$_\rho$ is independent on H and the in-plane ferromagnetic
behavior is suppressed below T$_\rho$. Descending of resistivity
due to in-plane ferromagnetism is killed and an upturn in
resistivity below T$_\rho$ is induced by magnetic field H $\perp$
ab plane, so that the resistivity shows similar behavior to that
of Na$_{0.5}$CoO$_2$\cite{Foo}. When H $\parallel$ ab plane,
T$_\rho$ increases with H and the charge-ordering behavior is
suppressed. Similar behavior is also observed in Na$_{0.5}$CoO$_2$
with high magnetic field as high as 45 T by Balicas \emph{et
al.}\cite{Balicas}. They found that the field H $\perp$ ab plane
enhances the charge-ordering state and the field H $\parallel$ ab
plane strongly suppresses the charge-ordering state.

The MR for H $\parallel$ ab plane and H $\perp$ ab plane in Fig.
2(c) and (d) shows different MR behavior. When H $\parallel$ ab
plane, the isothermal MR is negative and monotonously increases
with H. The magnitude of MR below T$_{\rho}$ is almost temperature
independent with varying H, and as high as 22$\%$ under H=13.5 T
at 20 K just below T$_{\rho}$. MR at 6 K is positive below 1 T and
changes to negative with increasing H as shown in the inset of
Fig. 2(d). The positive part at low field is considered to come
from antiferromagnetic background of large moments of
Co$^{3.5+\delta}$ sites, and the negative part from contribution
of in-plane ferromagnetic of small moments of Co$^{3.5-\delta}$
sites overwhelms the positive part at high field. When H $\perp$
ab plane, a crossover from negative to positive with increasing H
is observed in the isothermal MR. At 6 K, the MR is always
positive and as high as 12.5$\%$ under 13.5 T. Compared to
resistivity under magnetic field and MR of
Na$_{0.5}$CoO$_2$\cite{Wang1} and Na$_{0.55}$CoO$_2$\cite{Wang},
the results of resistivity under H and MR indicate that the
magnetic coupling of small moments of Co$^{3.5-\delta}$ sites can
be changed by magnetic field perpendicular to ab plane from
ferromagnetic to antiferromagnetic. An intriguing MR hysteresis
phenomenon is found below T$_\rho$ with increasing H to 14 T and
then decreasing to zero. The inset in Fig. 2(c) shows the
irreversible behavior. Below T$_\rho$, the MR hysteresis is
induced by H $\perp$ ab plane. The critical field (H$_C$) is
temperature dependent. Above T$_\rho$, the hysteresis phenomenon
is not observed with H as high as 14 T, indicating that the
hysteresis is related to in-plane ferromagnetism. In addition, the
direction of hysteresis varies with decreasing T. At 22 K and 25
K, the direction is clockwise. But the direction is anti-clockwise
 at 18 K. Such hysteresis is absent for H $\parallel$ ab plane.
 These phenomena can be explained with a spin-flop transition of small moments of
 $Co^{3.5-\delta}$ induced by the magnetic field perpendicular to ab plane.
The spin-flop transition makes the magnetic state from in-plane
ferromagnetic to antiferromagnetic state. The hysteresis is caused
by the different spin-flop energy of FM$\rightarrow$AF
(E$_{FM\rightarrow AF}$) and AF$\rightarrow$FM (E$_{AF\rightarrow
FM}$). The difference of hysteresis direction between 18 K and 22
K suggests that the relation of E$_{FM\rightarrow AF}$ and
E$_{AF\rightarrow FM}$ is changed with decreasing T.

The out-of-plane resistivity $\rho$$_c$ under H and the isothermal
out-of-plane MR are also studied. Fig. 3 shows the out-of-plane
magnetotransport and the isothermal out-of-plane MR with H
$\parallel$ ab plane and H $\perp$ ab plane, respectively. The
values of T$_\rho$ and T$_{c1}$ are the same as that of in-plane.
The effect of magnetic field on $\rho$$_c$ is quite similar to
that on $\rho$$_{ab}$ although the out-of-plane MR is less than
the in-plane MR. Similar to the in-plane behavior, T$_\rho$ is not
affected by H applied along c axis, while T$_\rho$ is enhanced by
H parallel to the ab plane. The charge-ordering behavior is
suppressed by H $\parallel$ ab plane and enhanced by H $\perp$ ab
plane. The behavior of isothermal out-of-plane MR is also similar
to that of the in-plane MR, and the MR hysteresis is also observed
only when H $\perp$ ab plane. These results further confirm the
spin-flop transition in this system.

\begin{figure}[h]
\centering
\includegraphics[width=9cm]{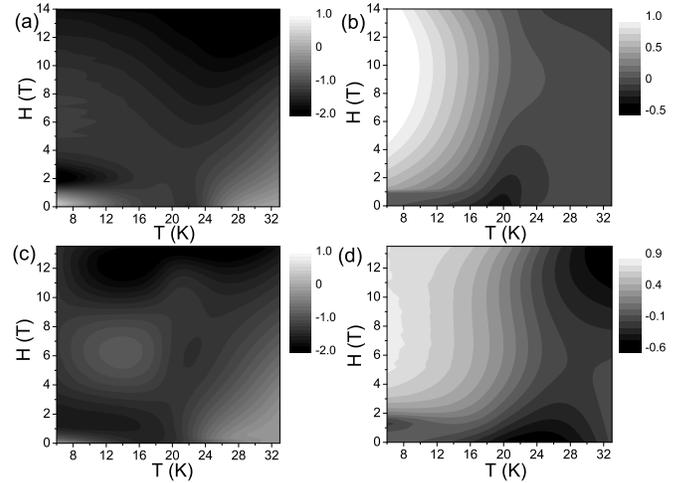}
\caption{2D project picture of derivation of MR dependent H and T.
(a): in-plane derivation of MR with H $\parallel$ Co-O plane; (b):
in-plane derivation of MR with H $\perp$ Co-O plane; (c):
out-of-plane derivation of MR with H $\parallel$ Co-O plane; (d):
out-of-plane derivation of MR with H $\perp$ Co-O plane.}
\label{fig4}
\end{figure}

The spin-flop transition for small moments of
 $Co^{3.5-\delta}$ leads to the metamagnetism. Such matamagnetism is difficult to
 detect by susceptibility measurements because small magnetic
moments order in the background of magnetic ordering with large
magnetic moments. To further understand the spin-flop transition,
a formula consisting of negative and positive contributions to MR,
-A$_{1}^{2}$ln(1+A$_{2}^{2}$H$^{2}$)+B$^{2}$H$^{n}$, is used to
qualitatively describe MR. The first term in the formula is
derived from the third-perturbation expansion of s-d exchange
Hamiltonian in the local magnetic moment model of
Toyozawa\cite{Khosla, Toyozawa}. The formula has been used to
explain the isothermal MR behavior for
Na$_{0.55}$CoO$_2$\cite{Wang} and another triangle lattice cobalt
oxide (Bi,Pb)$_2$Sr$_2$Co$_2$O$_y$\cite{Luo2}. Here, the negative
part is due to the contribution of in-plane magnetism ordered by
small moments of Co$^{3.5-\delta}$ sites. When the magnetic
coupling of small moments of Co$^{3.5-\delta}$ sites is changed
from ferromagnetic to antiferromagnetic due to spin-flop
transition induced by H $\perp$ ab plane, the contribution of
negative part to MR is changed. A 2D project picture of
$\frac{dMR}{dH}$ is plotted in Fig. 3. The in-plane
$\frac{dMR}{dH}$ with H $\parallel$ ab plane and H $\perp$ ab
plane is shown in Fig. 4(a) and (b), respectively. The black
region and white region stand for negative and positive slope
$\frac{dMR}{dH}$, respectively. It is found that the negative
maximum of the slope changes continuously for H $\parallel$ Co-O
plane and discontinuously for H $\perp$ ab plane with decreasing
temperature around the T$_\rho$. We believe that these differences
are due to metamagnetism induced by magnetic field perpendicular
ab plane. The derivation of the negative part is
-$\frac{2A_{1}^{2}A_{2}^{2}}{1+A_{2}^{2}H^{2}}$. When
H=$\frac{1}{A_{2}}$, the $\frac{dMR}{dH}$ reach to the maximum.
Khosla \emph{et al.}\cite{Khosla} gave the expression of A$_{2}$:
$\frac{\lambda}{T}$, $\lambda$=F(J). The parameter J is s-d
exchange integral. It suggests that the maximum of
$\frac{dMR}{dH}$ can be changed by variety of exchange integral.
Therefore, the discontinuity in $\frac{dMR}{dH}$ may come from the
change of magnetic coupling under s-d exchange model. It suggests
that the field H $\perp$ ab plane can induce antiferromagnetic
coupling of small moments of Co$^{3.5-\delta}$ sites below
T$_\rho$, which is corresponding to distinct change of the
exchange integral. The white region covers large area with H
$\perp$ ab plane and the transition point H$_c$
($\frac{dMR}{dH}$$|$$_{H_c}$=0) is dependent on temperature. No
apparent transition from negative slope to positive slope is
observed with H $\parallel$ ab plane. These results confirm that
the field H $\perp$ ab plane can induce antiferromagnetic coupling
and the field H $\parallel$ ab plane enhances the in-plane
ferromagnetism. The out-of-plane $\frac{dMR}{dH}$ is shown in Fig.
4(c) and (d), respectively. Similar behavior to the in-plane case
is observed for out-of-plane $\frac{dMR}{dH}$.

\begin{figure}[h]
\centering
\includegraphics[width=9cm]{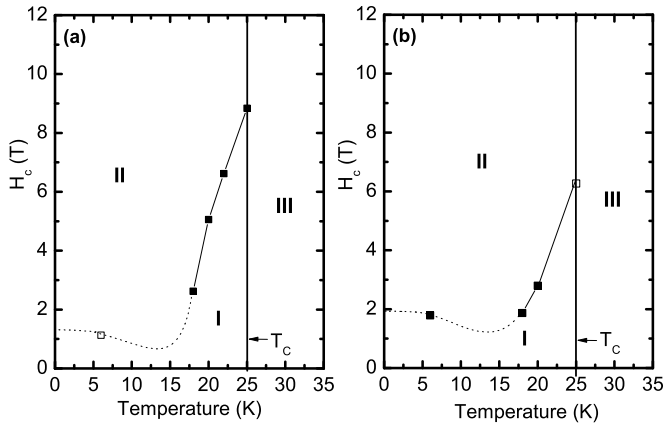}
\caption{Phase diagram for the spin of Co$^{3.5-\delta}$ in
Na$_{0.52}$CoO$_2$ with magnetic field perpendicular to ab plane
and temperature. (a): derived from in-plane data; (b): derived
from out-of-plane data. I region: in-plane ferromagnetic region;
II region: spin flop region; III region: paramagnetic region. Open
square is determined by the kink of $\frac{dMR}{dH}$ and filled
square is determined by $\frac{dMR}{dH}$=0 } \label{fig5}
\end{figure}

Based on the results that a spin-flop for small moments of
Co$^{3.5-\delta}$ sites is induced by H $\perp$ ab plane and the
in-plane ferromagnetism is suppressed, a H-T phase diagram with H
$\perp$ ab plane is derived as shown in Fig. 5. Three different
magnetic regions of small moments of Co$^{3.5-\delta}$ sites are
defined in H-T phase diagram: a) I: In-plane ferromagnetic region;
b) II: spin-flop region; c) III: paramagnetic region. The boundary
of I and II region is determined by $\frac{dMR}{dH}$=0. The
maximum of H$_c$ is 8.8 T for in-plane resistivity and 6.3 T for
out-of-plane resistivity, respectively. This phase diagram
indicates the instability of in-plane ferromagnetism with magnetic
field H $\perp$ ab plane. This instability suggests the existence
of novel magnetic ground state in this spin frustrated system
around x=0.5.

 The characteristics of spin-flop in Na$_x$CoO$_2$ have been
also reported for high Na content\cite{Bayrakci, Boothroyd,
Helme}. The neutron scattering studies on
Na$_{0.82}$CoO$_2$\cite{Bayrakci} and
Na$_{0.75}$CoO$_2$\cite{Boothroyd, Helme} indicate that the
in-plane and inter-plane spin correlations are ferromagnetic and
antiferromagnetic, respectively, and all the spins align along the
c axis. These results indicate that a spin-flop transition takes
place with decreasing Na content; that is, the spin direction
changes from along c axis to within ab plane. In
Na$_{0.85}$CoO$_2$, a magnetic field induced spin-flop has been
reported by Luo \emph{et al.}\cite{Luo}. Analogously, magnetic
field induced spin-flop transition of the small moments of
Co$^{3.5-\delta}$ sites occurs in our samples below T$_\rho$. When
H $\perp$ ab plane, the small moments of Co$^{3.5-\delta}$ sites
flop to c axis and the magnetic coupling changes from
ferromagnetic to antiferromagnetic. When H $\parallel$ ab plane,
the in-plane ferromagnetism is enhanced. As shown in Fig. 5, the
spin-flop transition occurs with decreasing temperature or
increasing magnetic field.

\begin{figure}[h]
\centering
\includegraphics[width=9cm]{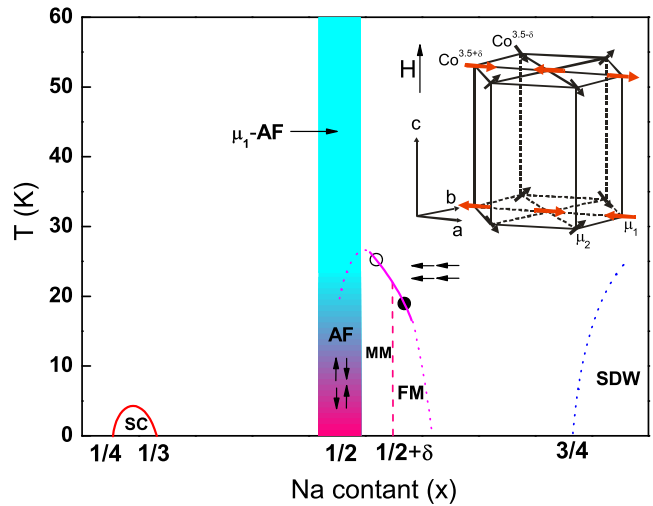}
\caption{The renewed magnetic phase diagram of Na$_x$CoO$_2$. The
$\mu_1$-AF, AF, FM, MM, SC and SDW stand Antiferromagnetism of
Co$^{3.5+\delta}$, Antiferromagnetism of Co$^{3.5-\delta}$,
Ferromagnetism of Co$^{3.5-\delta}$, Metamagnetism of
Co$^{3.5-\delta}$, Superconductivity and Spin density wave
respectively. The open and filled circles stand for the
ferromagnetic transition point for Na$_{0.52}$CoO$_2$ and
Na$_{0.55}$CoO$_2$, respectively. The right corner magnetic
structure is proposed for x=0.55 with field H $\perp$ ab plane.}
\label{fig6}
\end{figure}

Based on the results for x=0.5 and 0.55 reported by us\cite{Wang2,
Wang}, we renew the phase diagram around x=0.5 as shown in Fig. 6.
At x=0.5, the magnetic coupling of the small moments of
Co$^{3.5-\delta}$ sites are antiferromagnetic and the direction of
magnetic moments is along c axis. With increasing Na content, the
small moments of Co$^{3.5-\delta}$ sites flop to ab plane and the
magnetic coupling changes to ferromagnetic. For the samples with x
between x=0.5 and 0.55, magnetic field can induce spin-flop
transition. Because the spin-flop transition is accompanied with
the change of magnetic coupling, this region can be defined as
Metamagneitc region (MM). The magnetic structure for this region
with field H $\perp$ ab plane is shown in Fig.6. Many different
groups\cite{Sakurai, Milne, Sakurai2, Ihara} have revised the
superconducting phase. It is found that the superconductivity with
the highest T$_C$ is observed in the vicinity of a magnetic phase,
strongly suggesting that magnetic fluctuations play an important
role in the occurrence of superconductivity\cite{Sakurai2, Ihara}.
The valence of our sample is closed to +3.48 in the superconductor
Na$_{0.337}$(H$_3$O)$_{0.234}$CoO$_2$$\cdot$yH$_2$O. Therefore,
our result suggests that there exists strong correlation between
the superconductivity and novel magnetism around x=0.5. Recently,
a nesting scenario for charge-ordering Na$_{0.5}$CoO$_2$ has been
proposed by NMR and ARPES result\cite{Qian, Bobroff}. The nesting
leads to either spin density wave (SDW) or charge density wave at
low temperature which opens a gap on the nested parts of the fermi
surface. When the in-plane ferromagnetic is replaced by
antiferromagnetic coupling with field H $\perp$ ab plane, the
charge-ordering behavior is enhanced in Na$_{0.52}$CoO$_2$ as
shown in Fig. 1 and Fig. 2. Therefore, our result provides
indirect evidence to this nesting scenario. In addition, this
instability of fermi surface is considered to lead to
unconventional superconductivity.

In conclusion, the isothermal MR for the Na$_x$CoO$_2$ crystals
with x between 0.5 and 0.55 is systematically studied. An in-plane
ferromagnetism is observed below 25 K. When H $\perp$ ab plane,
the in-plane ferromagnetism is suppressed and the
antiferromagnetism of small moments of Co$^{3.5-\delta}$ sites is
enhanced, so that the resistivity shows similar behavior to that
observed in Na$_{0.5}$CoO$_2$. A striking feature is observed that
a MR hysteresis below the Curie temperature occurs with H $\perp$
ab plane. A magnetic field induced spin-flop transition of small
moments of Co$^{3.5-\delta}$ sites can explain these complex
phenomena well. Although the sample Na$_{0.52}$CoO$_2$ shows the
same in-plane ferromagnetism as that of Na$_{0.55}$CoO$_2$, their
resistivity under H and MR show quite different behavior. It
further indicates that the complicated magnetic structure around
x=0.5 is very sensitive to the Na content. Under the nesting
scenario, the instability in magnetism is related to the
instability of fermi surface, which is considered to lead to
unconventional superconductivity.

This work is supported by the grant from the Nature Science
Foundation of China and by the Ministry of Science and Technology
of China (973 project No: 2006CB601001 and 2006CB0L1205).

\end{document}